\documentclass[twocolumn,pra,showpacs]{revtex4}

\usepackage{graphicx}
\usepackage{dcolumn}
\usepackage{bm}

\preprint{APS/123-QED}

\begin{document}

\title{Photonic band gap via quantum coherence in vortex lattices of Bose gases}
\author{\"O. E. M\"ustecapl{\i}o\u{g}lu}
\affiliation{Ko\c{c} University, Department of Physics,
Rumelifeneri Yolu, 34450 Sar{\i}yer, Istanbul, Turkey}
\author{M. \"O. Oktel }
\affiliation{Bilkent University, Department of Physics, 06800 Bilkent, Ankara, Turkey}
\date{\today}
\begin{abstract}
We investigate the optical response of an  atomic Bose-Einstein
condensate with a vortex lattice. We find that it is possible for the vortex
lattice to act as a photonic crystal and create photonic band gaps, by enhancing the
refractive index of the condensate via a quantum coherent scheme.
If high enough index contrast between the vortex core and the
atomic sample is achieved, a photonic band gap arises depending on the healing length
and the lattice spacing. A wide range of experimentally accessible parameters are examined
and band gaps in the visible region of the electromagnetic spectrum are found. We also show
how directional band gaps can be used to directly measure the rotation frequency of
the condensate.
\end{abstract}
\maketitle

In-situ optical probing of vortices in rotating atomic
Bose-Einstein condensates (BECs) is a challenging task due to the
small size of vortices requiring imaging resolution close to a
single wavelength. It is usually necessary to let the condensate
expand ballistically before employing imaging methods
\cite{arv01,hce01,mcw00,mcb01,shk04}. In-situ imaging of the
vortex lattice has been demonstrated, by imaging the cloud
perpendicular to the axis of rotation and observing patterns
created by the overlapping lattice planes \cite{ech02}.
Alternative schemes based upon ultra-slow light imaging have been
proposed for non-destructive, in-situ dynamical imaging of
topological and collective excitations\cite{arto02} as well as for
probing phase evolution of the condensate\cite{ohbe02}.

In a simple treatment of light propagation in a vortex lattice of
a BEC, one may ignore the spatial periodicity in the system as the
vortex cores and the atomic sample would have negligible
difference in their refractive indices. The present paper however,
aims to exploit this periodicity to generate photonic band gaps,
similar to those in photonic crystals which are well-known for
their light localization effects\cite{pbg}. By modifying the
dispersion relation of light, photonic crystals create many
optical effects ranging from left-handed light propagation
\cite{a1} to slowing down or stopping light \cite{a2}. If the
vortex lattice of the condensate may be utilized as a photonic
crystal, these optical effects can be used to modify or probe the
condensate, or create novel atom-optical devices.

As a first example, presence of a photonic band gap may provide
stability to the vortex lattice against radiative losses
\cite{a3}. Such losses are especially problematic as a source of
decoherence in the quantum information applications, based upon
generation, storage and transfer of nonclassical states in
atom-optical systems\cite{dutt04,akam03,mair01,arna00,arlt99}.
These operations typically involve excited states and would suffer
from spontaneous emission. The idea of suppressing spontaneous emission
and enhancing stability of a Bose condensed system using photonýc band gaps
was first discussed for the case of optical lattices\cite{marz00}. Unfortunately,
due to the competing effects of weak excitations and the strong
periodic potential which are both provided by the same laser
beams forming the optical lattice, the suppression was found to be
insignificant\cite{marz00}. In the case of a vortex lattice, however,
the periodicity is a topological effect and the atoms can be
excited by a completely different and independent physical system
such as a pump laser. Hence, one would expect to achieve a more significant
suppression in the case of vortex lattices, which could be helpful
to improve performance of atom-optical devices. One may also
exploit the photonic band gap as an alternative to the existing
EIT schemes for slowing down or stopping  light, as discussed in a
number of recent papers \cite{a2}.

In this letter, we discuss presence of photonic band gaps in a
vortex lattice and show how these band gaps can be
used to measure the rotation frequency of a vortex lattice. In
vortex lattice experiments, the rotation rate of the vortex
lattice is not directly measured, but rather inferred from the
change of the cloud's aspect ratio upon
rotation\cite{arv01,hce01,mcw00,mcb01,shk04}. It is assumed that,
with a fully developed vortex lattice, the cloud rotates like a
solid body; on the other hand coupling of the collective modes of the
lattice with the cloud's collective oscillations may result in a
much more complicated behavior. An independent measurement of the
rotation frequency of the lattice would clarify the dependence of
the cloud aspect ratio on rotation as well as determine clearly
how far one can probe into the lowest Landau level regime
\cite{ho01}.

In order to have a significant photonic band gap, it is preferable
to have a material consisting of periodically alternating sequence
of dielectric domains with high contrast in their dielectric
constants. Dilute atomic gases exhibit strong differences from the
dielectric constant of free space only at their resonance
frequencies, where the optical fields would suffer high
absorbance. To overcome this challenge, we consider utilization of
a quantum coherent index enhancement scheme which allows for many
orders of magnitude improvement in the refractive index of an
atomic gas at frequencies where the absorption can be made zero\cite{scu91}.
This effect is based upon the similar principle of quantum coherence in
electromagnetically induced transparency (EIT)\cite{har97}, which is used
for slow-light\cite{liu01} and frozen light effects\cite{fle00}.

\begin{figure}
\centering{\vspace{0.5cm}}
\includegraphics[height=3.25in]{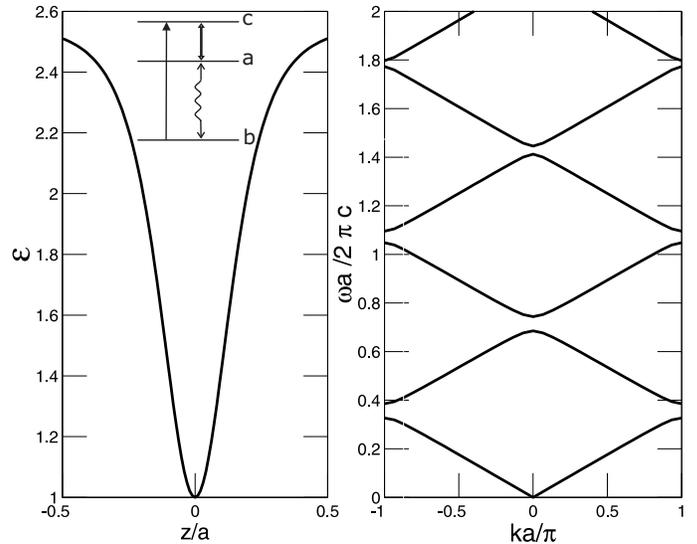}
\caption{(Left) Position dependence of the dielectric function of the one dimensional vortex lattice
within a spatial period where the background dielectric constant is assumed to be $\epsilon_0=1$.
Inset shows the level scheme of the upper-level microwave driven scheme.
(Right) Band structure of a one dimensional vortex lattice.}
\label{fig1}
\end{figure}

There are various schemes in which index enhancement accompanied
by vanishing absorption is possible, such as, the initial coherence
scheme, the Raman driven scheme, and the microwave driven
schemes\cite{scu97,fle92-2}. As an example, the upper-level microwave
scheme is illustrated with the energy level diagram in the inset
of Fig. \ref{fig1}. Quantum coherence is established through coupling
of levels $a$ and $c$ via a strong microwave field, which allows
for absorption free propagation of a probe field dipole coupled to
levels $a$ and $b$. A pump mechanism from level $b$ to $c$ ensures
an ultra high refractive index by maintaining a small fraction of
atoms in  level $c$. At the point of vanishing absorption,
susceptibility ($\chi$) becomes a real number and can be
estimated to be $\chi\sim
3N\lambda^3/40\pi^2$, where $N$ is the atomic density, and
$\lambda$ is the wavelength of $a-b$ transition. In a dense BEC,
the first correction to the susceptibility is argued to be the
local field correction\cite{localfield}, which changes
$\chi$ to $\chi/(1-\chi/3)$. Local field corrections are small
below $N\sim 10^{21}\mathrm{m}^{-3}$ at which $\chi$ increases to
$3.22$ from $1.55$ for $^{23}$Na gas of $\lambda = 589 \mathrm{nm}$.
For $^{87}$Rb gas of $\lambda = 794 \mathrm{nm}$,
$N=5.5\times 10^{20}\mathrm{m}^{-3}$, we find $\chi\sim 7$ including
the local field correction. The
dispersive behavior of $\epsilon$ near the frequency of index
enhancement will be ignored. This seems reasonable as
$\chi$ is slowly varying about its
peak value where the first derivative of $\chi$ vanishes
and the first correction is of second order in probe detuning.

As our atomic medium, we consider a vortex lattice of rotating BEC.
It has been predicted \cite{ho01,cst03,bay03,bro99},
and observed\cite{arv01,hce01,mcw00,mcb01,shk04,ces03} that the vortex lattice
is triangular. It has recently been shown that triangular lattice
is robust under variations of the trap potential in two
dimensions\cite{okte04}. The number of vortices can be as large as
few hundred. Presently, we assume the lattice is large and ignore
finite size effects. Lattice spacing $a$ can be compared to the
healing length $\xi = 1/\sqrt{8\pi N a_{sc}}$, where $a_{sc}$ is the scattering length.
We consider $a\sim 5-10\xi$. Our examinations indicate a rich variation of band structure
depending on $a$. Below, we report photonic band gaps in the visible range.

In the two-dimensional geometry, we consider straight line
vortices with perfect cylindrical symmetry for simplicity.
For vortices as effective cylinders of radius
$R=2\xi$, this
corresponds to a filling ratio $f=(2\pi/\sqrt{3})(R^2/a^2)\sim
14\%$. To study the effects of the vortex profile on the band
structure, we also consider a simpler system of one dimensional vortex
lattice. Such vortex row structures have theoretically been
predicted\cite{okte04,ssh04}, but have not been observed so far.
We use a vortex profile $\rho(z)$ in the Pad\'{e}
approximation\cite{ber03}. Our calculations show that
the form of vortex profile causes little change in
the band structure. Assuming a dilute
thermal gas background with uniform density,
dielectric constant within the vortex core region will be taken
as $\epsilon_0=1$. Defining $\epsilon_1=1+\chi$, dielectric function
of the atomic system can be expressed as $\epsilon(z) =
\epsilon_0+(\epsilon_1-\epsilon_0)\rho(z) = 1+\chi\rho(z)$.
We calculate the band structure of the vortex lattice of
$N = 6.6\times 10^{20}\mathrm{m}^{-3}$
$^{23}\mathrm{Na}$ atoms using plane wave expansion
method with $11\times 11$ grid of plane waves. Our results
for $a=10\xi$ are shown in Fig.\ref{fig1}. There is a
full gap at $\omega a/2\pi c=0.35$ with a gap-midgap ratio
of $0.143$. In a similar band structure for $^{87}$Rb, with
$N=5.5\times10^{20}\mathrm{m}^{-3}$, full gaps at visible midgap
frequencies $520$THz and $445$THz are found when
$a=10\xi$ and $a=5\xi$, respectively.

\begin{figure}
\centering{\vspace{0.5cm}}
\includegraphics[width=3.25in]{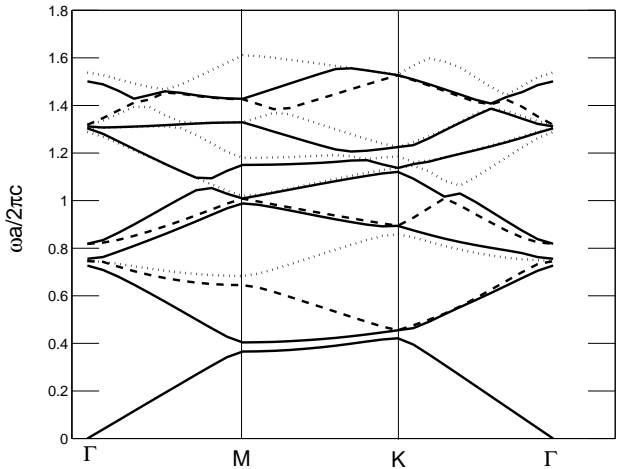}
\caption{TE band structure of a triangular vortex lattice with $a=10\xi$ of
$N=6.6\times 10^{20}\mathrm{m}^{-3}$ $^{23}$Na atoms. Different line patterns are
used for visual clarity.}
\label{fig2}
\end{figure}

Typical results for TE polarization
are shown in Fig.\ref{fig2} for the two-dimensional lattice.
Similar band structure occurs for
TM polarization. In the TM case, the gap
between the first and second bands decreases and becomes
negligibly small at the $K$ point of the Brillouin zone. For the
TE case, a gap appears at all points in the irreducible Brillouin
zone including the $M$ point. There are directional pseudogaps but
no complete band gap for both TM and TE polarizations. It should be emphasized
that it is actually possible to observe
full band gap, as it is typical for triangular lattice, for another
choice of parameters. Assuming $a>\sim 4 \xi$ gives $R/a<\sim 0.5$
for the radius of the vortex core $R=2\xi$. A full
gap for TE polarization is found for $^{87}$Rb with
$a=5\xi$ at $N=5.5\times 10^{20} \mathrm{m}^{-3}$. Full gaps for both polarizations
may appear at higher index-contrasts.

After investigating an infinite vortex lattice and finding that band gaps develop,
we consider whether the effects discussed before can be observed in an experiment
where the vortex lattice is necessarily finite. In a typical experiment, there are three major
differences from the idealized vortex lattice considered above. Here, we consider
all three separately.

\begin{figure}
\centering{\vspace{0.5cm}}
\includegraphics[width=3.25in]{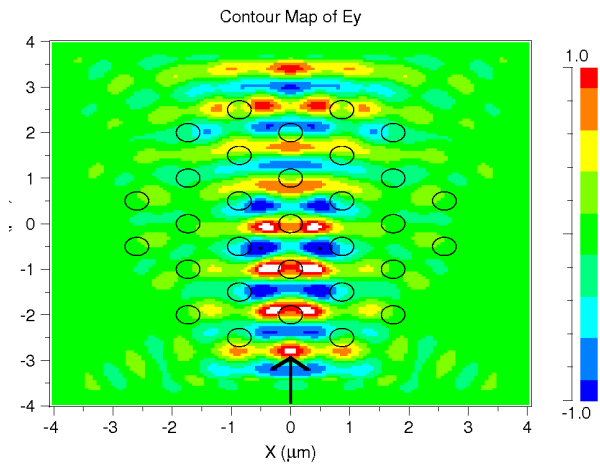}
\caption{(Color online) A vortex lattice with 31 vortices.
Vortex spacing is taken to be 5 times the vortex core size which is
equal to the light wavelength. Vortex cores have $n=1$ while BEC has index of refraction $n=3$.
With these parameters
infinite lattice results indicate a directional band gap, which can be observed by
comparing this figure with the
Fig.\ref{fig4}. Direction of the incident light is denoted by an arrow in both figures.}
\label{fig3}
\end{figure}

\begin{figure}
\centering{\vspace{0.5cm}}
\includegraphics[width=3.25in]{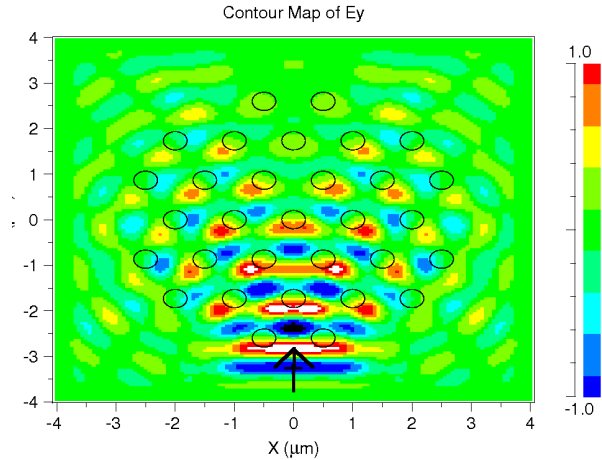}
\caption{(Color online) Same finite vortex lattice as in Fig.\ref{fig3}, but rotated by $\pi/6$.
A directional band gap is clearly seen as there is no transmission.}
\label{fig4}
\end{figure}

First, we assumed a strictly two dimensional vortex lattice, which disregards any bending or motion of the vortices along
the rotation axis. However, the bending of the vortices have theoretically been predicted to be small except near the edges
of the cloud \cite{ho01,chst03}. This prediction was later confirmed by viewing the vortex lattice from the side
\cite{ech02}. As the deviation of the
position of the vortices are much less than the vortex spacing, we expect the bending, or Kelvin mode  excitations
of the vortices in the third dimension to be a very small effect.

Second simplification in our calculation was that the index profile of the cloud remained uniform
except for the appearance of vortices. In a trapped BEC with vortices overall density profile is maximum in the middle
and would slowly decay towards the edges. Average density profile is predicted to be Gaussian \cite{ho01}
 or Thomas-Fermi like \cite{wbp04}, in both
cases a monotonous slowly varying function. This slow variation of the index profile would create an extra source of
scattering from the cloud, especially under coherent index enhancement. The presence of this extra scattering may complicate
the optical band structure effects if the variation of the density is not slow on the scale of vortex spacing. However, as the
overall density decays monotonously, a fast variation of the density on the scale of vortex spacing would mean there are only
a small number of vortices in the cloud. So for a BEC containing enough number of vortices to create band structure effects,
photonic band structure should be superimposed on the extra scattering effect which is expected to vary slowly in frequency.
Thus, the photonic crystal effects should be easily observable.

A third, and related to second, concern in an experiment would be the finiteness of the vortex lattice. We performed a
numerical simulation of the propagation of an electromagnetic wave in a finite lattice to quantify the number of vortices
needed to observe band structure effects. In Figs.\ref{fig3}-\ref{fig4}, we display our results for 31 vortices in two different
orientations. The simulation was done with parameters where the infinite lattice  calculation indicates a directional band gap, and this
band gap is clearly seen by contrasting the two figures. In Fig.\ref{fig3}, a Gaussian beam passes through the vortex lattice without
much scattering, while in Fig.\ref{fig4}, after a rotation of the vortex lattice by $\pi/6$,  there is no transmission.
Our results indicate that around 20 vortices are enough to qualitatively observe the infinite lattice predictions for the
band gaps.

An important consequence of such a directional band gap is that it can be utilized to measure the rotation frequency of
the BEC. In the vortex lattice experiments,
the rotation frequency
of the vortex lattice is measured indirectly by inferring it from the radius of the cloud \cite{arv01,hce01,mcw00,mcb01,shk04,ech02}.
However, if a directional band gap is established, probing laser beam will be
chopped by the rotation of the vortex lattice and counting the chopping frequency will yield a direct measurement of the
rotation frequency. Such a measurement can be carried out non-destructively, and continuously. As the rotation frequency
of the cloud, which is around hundred Hertz in a typical experiment, is much smaller than the laser frequency, rotation
will not affect the optical band structure discussed earlier.

We have shown that photonic band gaps can be generated in vortex lattices of an atomic
Bose-Einstein condensate by utilizing quantum coherence based ultra-high index enhancement effect.
We considered a two dimensional triangular lattice and found the physical parameter
regime for band gaps and other band structure effects to be observed. We argued that the effects found
for an infinite lattice could be observed in experiments where tens of vortices are created in the condensate.
We finally suggested a method to use a directional band gap to measure the rotation frequency of the condensate
non-destructively.

\"O.E.M. acknowledges support from TUBA-GEBIP Award. M.\"O.O. acknowledges support from
TUBITAK-KARIYER grant, and thanks E. \"Ozbay
for computational support.


\begin{thebibliography}{}

\bibitem{arv01}
J.~R.~Abo-Shaeer, C.~Raman, J.~M.~Vogels and W.~Ketterle,
Science {\bf 292}, 476 (2001).

\bibitem{hce01}
P.~C.~Haljan, I.~Coddington, P.~Engels and E.~A.~Cornell,
Phys. Rev. Lett. {\bf 87}, 210403 (2001).

\bibitem{mcw00}
K.~W.~Madison, F.~Chevy, W.~Wohlleben, and J.~Dalibard,
Phys. Rev. Lett. {\bf 84}, 806 (2000).

\bibitem{mcb01}
K.~W.~Madison, F.~Chevy, V.~Bretin, and J.~Dalibard,
Phys. Rev. Lett. {\bf 86}, 4443 (2001).

\bibitem{shk04}
N.~L.~Smith, W.~H.~Heathcote, J.~M.~Krueger, and C.~J.~Foot,
Phys. Rev. Lett. {\bf 93}, 080406 (2004).

\bibitem{ech02}
P.~Engels, I.~Coddington, P.~C.~Haljan and E.~A.~Cornell,
Phys. Rev. Lett. {\bf 89}, 100403 (2002).

\bibitem{arto02}M. Artoni and I. Carusotto, Phys. Rev. A {\bf 67}, 011602(R) (2003).

\bibitem{ohbe02} P. \"Ohberg,
Phys. Rev. A {\bf 66}, 021603(R) (2002).

\bibitem{pbg} E.~Yablonovitch, T.J. Gmitter, and K.M. Leung,
Phys. Rev. Lett. {\bf 67}, 2295 (1991); A. Mekis, S. Fan, and J.D.
Joannopoulos, Phys. Rev. B {\bf 58}, 4809 (1998).

\bibitem{a1}
E.~\c{C}ubuk\c{c}u, K.~Ayd{\i}n, E.~\"Ozbay, S.~Foteinopoulou, and C.~M.~Soukoulis,
Nature(London) {\bf 423}, 604 (2003).

\bibitem{a2}
M.~F.~Yan{\i}k, and S.~Fan, Phys. Rev. Lett. {\bf 92}, 083901
(2004); L.D. Negro {\it et al.}, {\it ibid.} {\bf 90}, 055501
(2003).

\bibitem{a3}
E.~Yablonovitch, Phys. Rev. Lett. {\bf 58}, 2059 (1987).

\bibitem{dutt04} Z. Dutton and J. Ruostekoski, Phys. Rev. Lett.
{\bf 93} 193602 (2004).

\bibitem{akam03} D. Akamatsu and M. Kozuma,
Phys. Rev. A {\bf 67}, 023803 (2003).
%
%
\bibitem{mair01} A. Mair, A. Vaziri, G. Weihs, and A. Zeilinger,
Nature (London) {\bf 412}, 313 (2001).
%
\bibitem{arna00} H.H. Arnaut and G.A. Barbosa,
Phys. Rev. Lett. {\bf 85}, 286 (2000).
%
\bibitem{arlt99} J. Arlt, K. Dholakia, L. Allen, and M.J. Padgett,
Phys. Rev. A {\bf 59}, 3950 (1999).
%
\bibitem{marz00} K.-P. Marzlin and W. Zhang,
Eur. Phys. J. D {\bf 12}, 241 (2000).

\bibitem{ho01}
T-L.~Ho,
Phys. Rev. Lett. {\bf 87}, 060403 (2001).

\bibitem{scu91}
M.O. Scully, Phys. Rev. Lett. {\bf 67}, 1855 (1991).

\bibitem{har97}
S.E. Harris, Physics Today {\bf 50}, 36 (1997), and references therein.

\bibitem{liu01}
C. Liu, Z. Datton, C.H. Behroozi, L.V. Hau,
Nature {\bf 409}, 490 (2001).

\bibitem{fle00}
M. Fleischhauer and M.D. Lukin,
Phys. Rev. Lett.  {\bf 84}, 5094 (2000).

\bibitem{scu97}
M.O.~Scully and M.S.~Zubairy, {\it Quantum Optics},
(Cambridge University Press, Cambridge, 1997).

\bibitem{fle92-2}
M. Fleischhauer, C.H. Keitel, M.O. Scully, C. Su, B.T. Ulrich, and S.-Y. Zhu
Phys. Rev. A {\bf 46}, 1468 (1992).

\bibitem{localfield} M. Fleischhauer, Phys. Rev. A {\bf 60}, 2534
(1999); H. Wallis {\it ibid.} {\bf 56}, 2060 (1997); O. Morice
{\it et al.}, {\it ibid.} {\bf 51}, 3896 (1995); J. Ruestekoski
and J. Javanainen, {\it ibid.} {\bf 56}, 2056 (1997); K.V.
Krutitsky, K.-P. Marzlin, and J. Audretsch, {\it ibid.} {\bf 65},
063609 (2002).


\bibitem{cst03}
M.~Cozzini and S.~Stringari,
Phys. Rev. A {\bf 67}, 041602(R) (2003).

\bibitem{bay03}
G.~Baym, Phys.
Rev. Lett. {\bf 91}, 110402 (2003).

\bibitem{bro99}
D.~A.~Butts and D.~S.~Rokshar,
Nature(London) {\bf 397}, 327 (1999).

\bibitem{ces03}
I.~Coddington,P.~Engels,V.~Schweikhard and E.~A.~Cornell,
Phys. Rev. Lett. {\bf 91}, 100402 (2003).

\bibitem{okte04} M.\"O. Oktel,
Phys. Rev. A {\bf 69}, 023618 (2004).

\bibitem{ssh04} S.~Sinha, and G.~Shylapnikov cond-mat/04112139

\bibitem{ber03} N.G. Berloff, Jour. Phys. A {\bf 37}, 1617 (2004).

\bibitem{chst03}
F.~Chevy and S.~Stringari,
Phys. Rev. A {\bf 68},053601 (2003).

\bibitem{wbp04}
G.~Watanabe, G.~Baym, and C.~J.~Pethick, Phys.
Rev. Lett. {\bf 93}, 190401 (2004).

\end{thebibliography}
\end{document}